\documentclass[12pt]{iopart}

\usepackage{graphicx}
\usepackage{iopams}
\usepackage{revsymb}

\begin{document}

\title[ S Longhi and A Szameit, Dynamic localization in ...]{Dynamic localization in Glauber-Fock lattices}

\author{S Longhi}

\address{Dipartimento di Fisica, Politecnico di Milano, Piazza L. da Vinci 32, I-20133 Milano,
Italy}
\ead{longhi@fisi.polimi.it}

\author{A Szameit}

\address{Institute of Applied Physics, Abbe Center of Photonics, Friedrich-Schiller-Universit\"{a}t Jena, Max-Wien-Platz 1, 07743 Jena,
Germany}

\begin{abstract}
\noindent

Glauber-Fock lattices refer to a special class of semi-infinite tight-binding lattices with inhomogeneous hopping rates which are found  in certain simple solid-state, quantum optics and quantum field theoretical models. Here it is shown that dynamic localization, i.e. suppression of quantum diffusion  and periodic quantum self-imaging by an external sinusoidal force [D.H. Dunlap and V.M. Kenkre, Phys. Rev. B {\bf 34}, 3625 (1986)], can be exactly realized in Glauber-Fock lattices, in spite of inhomogeneity of hopping rates and lattice truncation. 

\end{abstract}

\pacs{73.23.-b, 73.63.-b, 73.40.Gk, 42.82.Et}


\maketitle

\section{Introduction}
Dynamic localization (DL), the suppression of quantum diffusion in tight-binding lattices by a sinusoidal force, was suggested more than 20 years ago by Dunlap and Kenkre \cite{Dunlap}. For appropriate driving conditions, they found that any initial distribution of the particle occupation probability is restored  after each modulation cycle of the applied force, realizing periodic quantum self-imaging and thus suppression of quantum diffusion.  Earlier studies in this field were focused on semiconductor superlattices  in strong THz fields, and extensions of the original theory were proposed to account for non-sinusoidal fields, combined ac and dc driving forces,  non-nearest neighboring couplings, and particle interactions (see, for instance, \cite{T1a,T1b,T1c,T2a,T2b,T3a,T3b,T4,T5,T6}). In the framework of Floquet theory for time-periodic quantum Hamiltonians \cite{F1,Grifoni}, Holthaus showed that DL is related to the collapse of the quasi-energy minibands of the superlattice under certain driving conditions. Demonstrations of DL have been  reported in a series of recent experiments using ultracold atoms in driven optical lattices \cite{Ari1,Ari2,Ari3,Ari4} and optical beams in engineered photonic lattices \cite{OP1,OP2,OP3,OP4,OP5,OP6,OP7}. Unfortunately, truncation effects or inhomogeneities in the lattice generally prevent DL and exact self-imaging  \cite{T1b,T1c,uffa,Villas,Longhi08,Garanovich1,Garanovich2}. The role of local defects and disorder in tight-binding systems in the presence of a strong external ac electric field has been investigated in several early works on  DL (see, for instance, \cite{T1c,refer1,refer2,refer3}). For local defects, it was found that the appearance and disappearance of localized states, as well as their localization lengths, can be controlled by the amplitude or frequency of the driving field \cite{T1c}. Random  inhomogeneities generally lead to an imperfect quasi-energy  collapse, however a crossover from merely DL to Anderson-like localization is found in this case \cite{refer1,refer2}.
In finite  chains of coupled wells with uniform hopping rates the quasi-energy collapse is imperfect, and a pseudo-collpase is typically observed with a fine and generally complex structure of crossings and anticrossings of quasi-energies near the pseudo collapse point \cite{Villas,Longhi08}. In a homogeneous semi-infinite lattice, the application of the sinusoidal force can give rise to dynamically-sustained surface states \cite{Garanovich1,Garanovich2,GaranovichReview}, which prevent quantum self imaging when the wave packet reaches the lattice boundary. As a general rule, DL in truncated or inhomogeneous lattices is only asymptotically realized in the high frequency regime, where  the applied force basically yields a renormalization of the tunneling rate and the DL condition corresponds to the coherent destruction of tunneling between adjacent sites in the lattice  \cite{Grossmann}.  A natural question arises whether there exist truncated or inhomogeneous tight-binding lattices in which the application of a sinusoidal force leads to DL and quantum self-imaging like in homogeneous and infinitely-extended lattices, even for low-frequency driving forces. A special class of inhomogeneous semi-infinite lattices is represented by the so-called Glauber-Fock lattices \cite{SzamGF1,SzamGF2,SzamGFb}. Such lattices have been recently introduced to simulate classical analogues of displaced Fock
(number) states  and realized using evanescently-coupled optical waveguides with engineered coupling constants \cite{SzamGF1,SzamGF2}. They are also found in the Fock space representation of certain solid-state models or simple quantum field theoretical models \cite{QF1,QF2,QF2a,QF3,QF4a,QF4b}, such as the atomic limit of the Hubbard-Holstein model describing electron-phonon interaction in a two-site potential \cite{QF4a,QF4b} or the deep strong coupling regime 
of circuit or cavity quantum electrodynamics with degenerate qubit energy levels \cite{QF2,QF2a}. In the optical context, Glauber-Fock waveguide lattices have been shown to support a new family of intriguing quantum correlations not encountered in uniform lattices, which can be probed by excitation with non-classical states of light \cite{SzamGF2,SzamGFb}. From this perspective, such lattices could therefore provide a fertile ground to test and engineer quantum random walks of correlated particles \cite{SzamGFb}. The additional degree of freedom provided by the ac driving field could be exploited to control quantum correlations and the walk in the lattice.\par
In this work  we show theoretically that Glauber-Fock lattices represent an important class of inhomogeneous and truncated tight-binding lattice models where DL and quantum self-imaging can be exactly realized. This result basically arises from the fact that the Glauber-Fock lattice dynamics maps in Fock space the evolution of a bosonic field with a time-periodic Hamiltonian, which admits of a periodic dynamics.\\  

\section{Ac-driven Glauber-Fock lattices: model and dynamic localization}
The coherent hopping motion of a quantum particle in a semi-infinite one-dimensional chain of coupled wells with inhomogeneous hopping rates driven by an external spatially-homogeneous and time-dependent force $F(t)$ [see Fig.1(a)] is described by a set of  tight-binding coupled equations for the amplitude occupation probabilities $c_n(t)$ of the $n$-th lattice site \cite{Dunlap}
\begin{equation}
i \frac{dc_n}{dt}=-\kappa_{n+1} c_{n+1}-\kappa_{n}c_{n-1}+nF(t)c_n
\end{equation}
where $\kappa_n$ is the hopping rate between nearest-neighbor lattice sites $n$ and $(n-1)$, $n=0,1,2,3...$, and $\kappa_0=0$. For a Glauber-Fock lattice one has
\begin{equation}
\kappa_n=\rho \sqrt{n}
\end{equation}
with $\rho$ constant \cite{note0}. After the formal substitution $c_n(t)=b_n(t) \exp[-in \Phi(t)]$, Eqs.(1) can be written in the equivalent form
\begin{equation}
i \frac{db_n}{dt}=-\kappa_{n+1}(t) b_{n+1}-\kappa^*_{n}(t)b_{n-1}
\end{equation}
where we have set
\begin{equation}
\Phi(t)=\int_0^t dt'F(t') 
\end{equation}
and
\begin{equation}
\kappa_n(t)=\kappa_n \exp[-i \Phi(t)].
\end{equation}
\begin{figure}
 \includegraphics[scale=0.7]{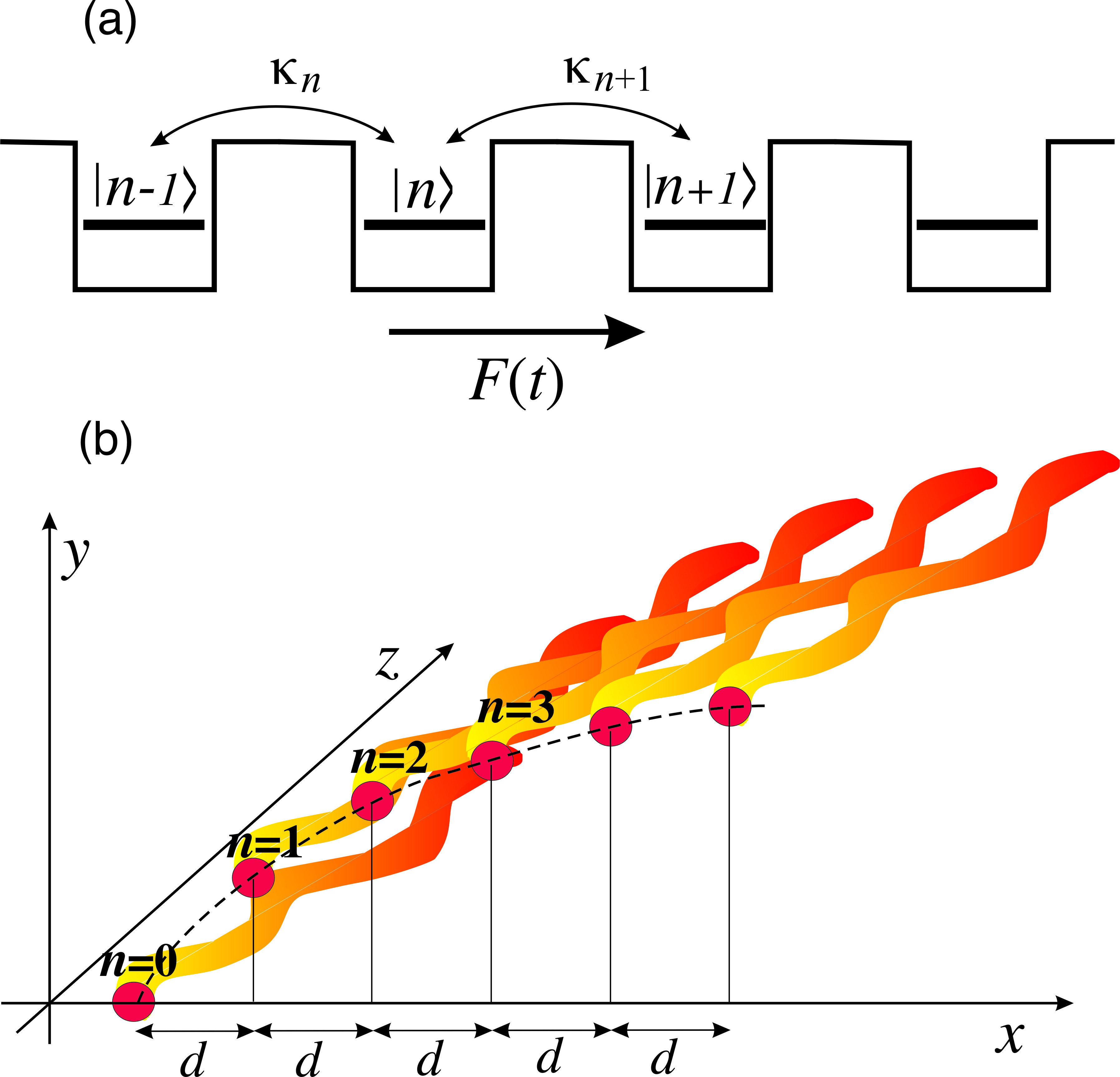}
\caption{(Color online) (a) Schematic of an ac-driven tight-binding lattice with inhomogeneous hopping rates, and (b) a possible experimental implementation based on light transport in a chain of evanescently-coupled optical waveguides with a periodically-curved optical axis.}
\end{figure}
 As shown e.g. in Ref.\cite{Longhi08}, inhomogeneities in the hopping rates and lattice truncation generally prevent exact DL and quantum self-imaging. On the other hand, for a time-independent external force quantum self-imaging, corresponding to generalized Bloch oscillations, have been found in a few types of lattices with inhomogeneous hopping rates or defects (see, for instance, \cite{LonghiPRBR,Longhi1a,Longhi1b}). The Glauber-Fock lattices with a time-independent force belong to such a class of special lattices. Indeed, in the time-independent case it is known that exact diagonalization of the lattice Hamiltonian is possible via  Lang-Firsov transformation \cite{Lang}, and the energy spectrum is given by an equally-spaced Wannier-Stark ladder spectrum (see, for instance, \cite{QF2}). Bloch-like oscillations in Glauber-Fock lattices driven by a time-independent force have been recently demonstrated in photonic model systems based on light transport in arrays of evanescently-coupled optical waveguides with engineered coupling constants \cite{QF2a,SzamOL}. In Ref.\cite{SzamGFb} it was also shown that periodic revivals are also possible for a constant force $F(t)=F_0$ and for a period modulation of the (real-valued) coupling constant $\rho(t)$, provided that  a resonance condition between the external force and the modulation frequency of the hopping rate is met. Here we consider the case of a general time-dependent force with a constant hopping rate $\rho$ \cite{note0}, and prove that exact DL and quantum self-imaging can be realized in Glauber-Fock lattices like in a homogeneous and infinitely-extended lattice. To this aim, let us  notice that Eqs.(3) describe the evolution in Fock space  of the following time-dependent Hamiltonian for a bosonic field
 \begin{equation} 
 \hat{H}(t)=-\rho(t) \hat{a}-\rho^*(t) \hat{a}^{\dag}
\end{equation}
where 
\begin{equation}
\rho(t)= \rho \exp [-i \Phi (t)]
\end{equation}
and $\hat{a}$, $\hat{a}^{\dag}$ satisfy the usual commutation relations of bosonic operators.  In fact, if the state vector $|\psi(t) \rangle$ of the bosonic field is decomposed in Fock space as $|\psi(t) \rangle =\sum_{n=0}^{\infty} b_n(t) (1/ \sqrt{n !}) \hat{a}^{\dag n} |0 \rangle \equiv \sum_{n=0}^{\infty} b_n(t) | n \rangle$, the evolution equations for the amplitude probabilities $b_n(t)$ are precisely given by Eqs.(3). Hence the solution to Eqs.(3) is formally given by
\begin{eqnarray}
b_n(t) & = & \langle n | \psi(t) \rangle =\langle n | \hat{U}(t) | \psi(0) \rangle \nonumber \\
& = & \sum_{l=0}^{\infty} \langle n | \hat{U}(t) | l  \rangle  b_l(0)
\end{eqnarray}
where the evolution operator (propagator) $\hat{U}(t)$ satisfies the Schr\"{o}dinger equation (with $\hbar=1$)
\begin{equation}
i \frac{d \hat{U}}{dt}= \hat{H}(t) \hat{U}(t)
\end{equation}
with $\hat{U}(0)=\hat{I}$ (the identity operator).
Note that, since the Hamiltonian $\hat{H}(t)$ is time-dependent and in general $\hat{H}(t)$ does not commute with $\hat{H}(t')$ for $t' \neq t$, the propagator 
$\hat{U}(t)$ has to be computed as a time-ordered exponential, and thus the relation $\hat{U}(t) = \exp[-i \int_0^t dt' \hat{H}(t')]$ is generally wrong, except for some special cases. Indeed,  let us suppose that, for any $t$, the following commutation relation holds
\begin{equation}
\left[ \int_0^t dt' \hat{H}(t') , \hat{H} (t) \right]= \alpha(t) \hat{I}
\end{equation}
where $\alpha(t)$ is an arbitrary function of time. Then, by use of the Baker-Campbell-Hausdorff formula it can be readily proven that one has
\begin{equation}
\hat{U}(t)=G(t) \exp \left[   -i \int_0^t dt' \hat{H}(t') \right]
\end{equation}
where the function $G(t)$ is given by
\begin{equation}
G(t)=\exp \left[  -\frac{1}{2} \int_0^t dt' \alpha (t') \right].
\end{equation}
For the Hamiltonian Eq.(6) associated to the driven Glauber-Fock lattice, one simply has $\int_0^t dt' \hat{H}(t')= -\sigma(t) \hat{a}-\sigma^*(t) \hat{a}^{\dag}$, where
\begin{equation}
\sigma (t) \equiv \rho \int_{0}^{t} dt' \exp[-i \Phi(t')].
\end{equation}
Hence
\begin{eqnarray}
\left[ \int_0^t dt' \hat{H}(t'), \hat{H}(t)  \right] & = & 2 i  \; {\rm Im} \left\{ \sigma(t) \rho^*(t) \right\} [\hat{a},\hat{a}^{\dag}] \nonumber \\
& = & 2 i \; {\rm Im} \left\{ \sigma(t) \rho^*(t) \right\},
\end{eqnarray}
i.e. the condition (10) is satisfied, with $\alpha(t)=2 i \; {\rm Im} \left\{ \sigma(t) \rho^*(t) \right\}$. According to Eq.(11), the evolution operator for the Hamiltonian (3) reads explicitly
\begin{eqnarray}
\hat{U}(t) & = & \exp \left[ -i \int_0^t dt' \;  {\rm Im} \left\{   \sigma(t') \rho^*(t') \right\} \right] \nonumber \\
& \times &  \exp \left[ -i \sigma(t) \hat{a} - i \sigma^*(t) \hat{a}^{\dag} \right]
\end{eqnarray}
where $\rho(t)$ and $\sigma(t)$ are defined by Eqs.(7) and (13), respectively.
Equation (15) shows that, apart from an inessential phase term, the evolution of the driven Glauber-Fock lattice is basically obtained from the undriven lattice dynamics by the replacement $\rho t \rightarrow \sigma(t)$. Let us assume now that at a time $t=T$ one has $\sigma(T)=0$, i.e. that
\begin{equation}
\int_{0}^{T} dt \exp [-i \Phi(t)]=0
\end{equation}
 According to Eq.(15), $\hat{U}(t=T)$ is the identity operator, apart from an inessential phase term, and thus self-imaging is attained at $t=T$.  For an ac force $F(t)=F(t+T)$ of period $T= 2 \pi / \omega$, Eq.(16) is precisely the condition for DL found in a uniform and infinitely-extended lattice (within the single-band and nearest-neighbor tight-binding approximations), i.e. the Glauber-Fock lattice behaves in the presence of the external ac force like a uniform and infinitely-extended lattice. In particular, for a sinusoidal driving force $F(t)=F_0 \cos (\omega t)$, Eq.(16) yields the well-known DL condition originally found by Dunlap and Kenkre \cite{Dunlap}
 \begin{equation}
 J_0 \left( \frac{F_0}{\omega} \right)=0
 \end{equation}
 \begin{figure}
 \includegraphics[scale=0.4]{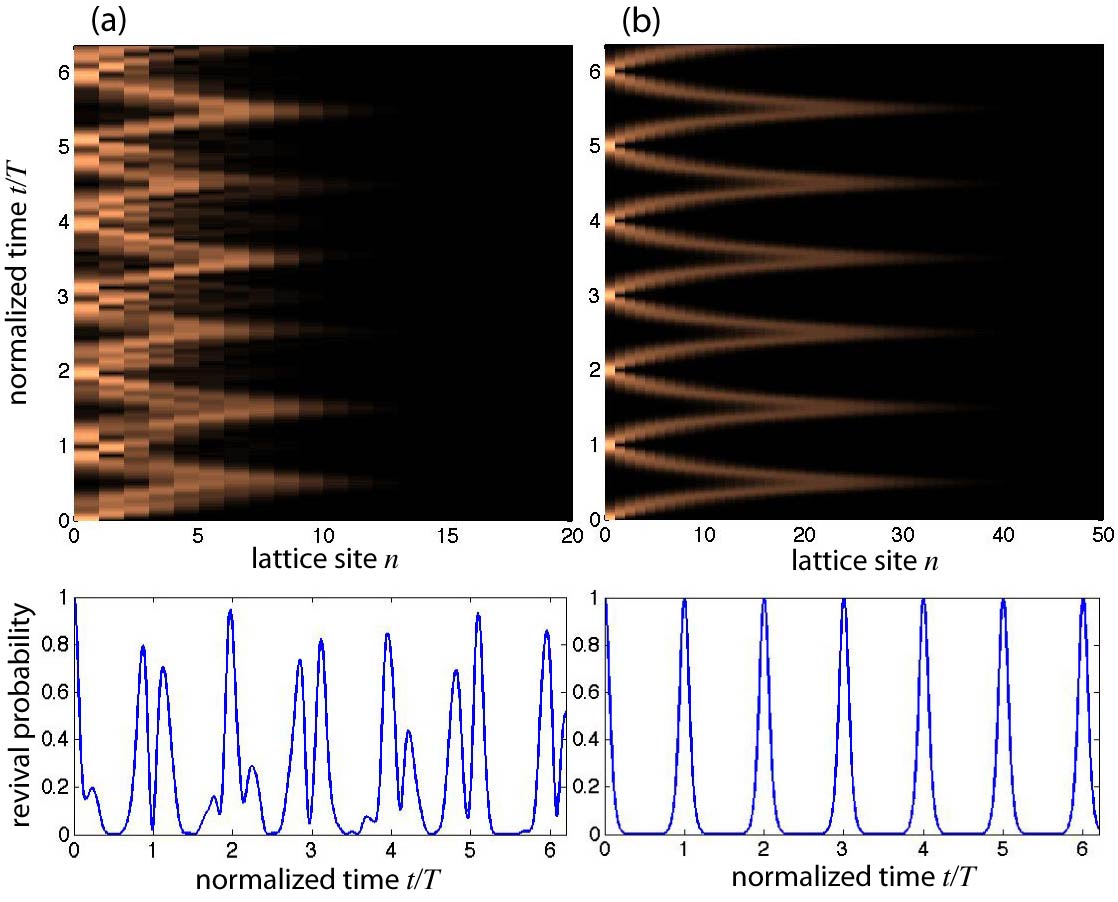}
\caption{(Color online) Evolution of the amplitude probabilities $|c_n(t)|$ (upper plots) and of the revival probability $P_r(t)$ (lower plots) of a quantum particle driven by a sinusoidal force in (a) a semi-infinite lattice with homogeneous hopping rate, and (b) in a Glauber-Fock lattice. The initial excitation condition is $c_{n}(0)=\delta_{n,0}$. Parameter values are given in the text.}
\end{figure}
 \begin{figure}
 \includegraphics[scale=0.4]{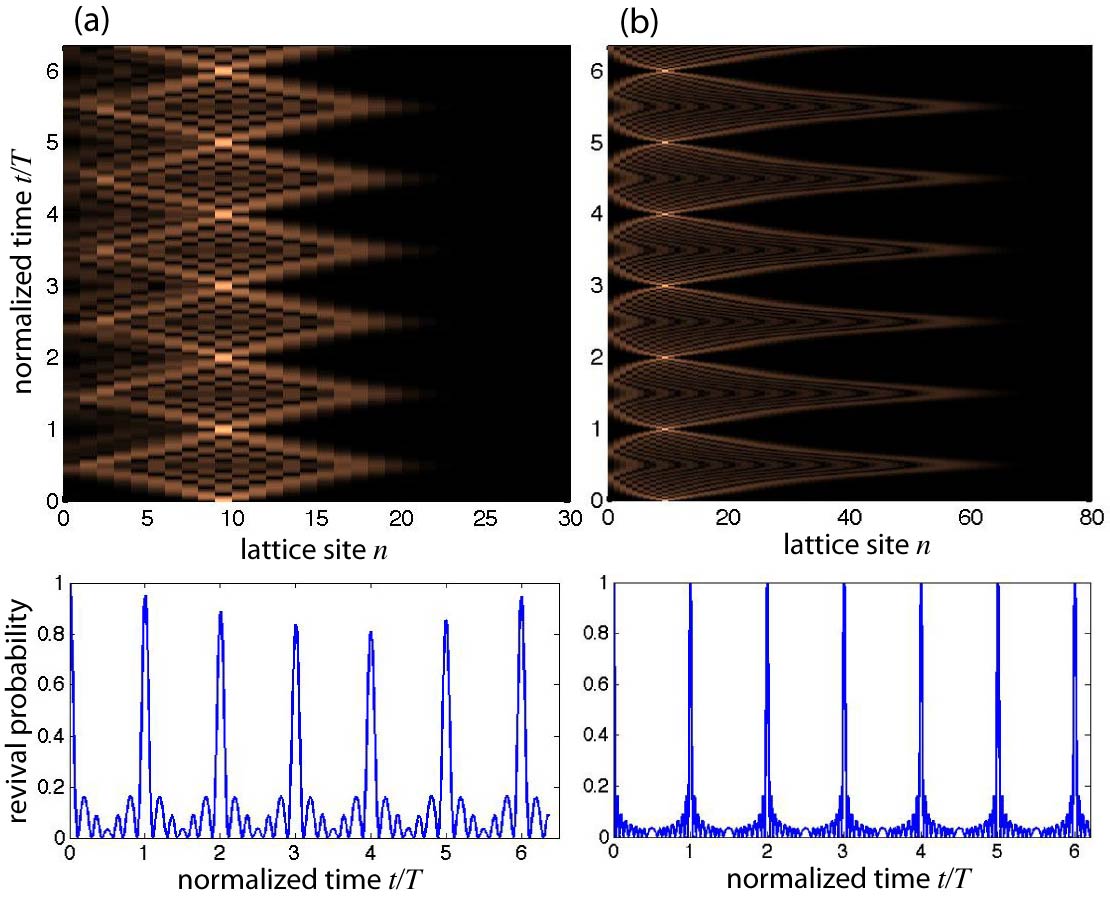}
\caption{(Color online) Same as Fig.2, but for the different initial excitation condition $c_n(0)=\delta_{n,10}$.}
\end{figure}
where $J_0$ is the Bessel function of first kind and zero order. It should be noted that such a result holds regardless of the ratio between the hopping rate $\rho$ and the modulation frequency $\omega$, i.e. it is valid even for arbitrarily small modulation frequencies \cite{note}. Also, the proof provided above does not require the knowledge  of the exact  impulse response of  the lattice.

\section{Numerical simulations and photonic realization}

We have checked the predictions of the theoretical analysis by direct numerical simulations of Eqs.(1). As an example, Figs. 2 and 3 show the numerically-computed evolution of the lattice site occupation amplitudes $|c_n(t)|$ and of the return (revival) probability $P_r(t)=|\langle \psi(t)| \psi(0) \rangle |^2$ for the two different initial excitations $c_{n,0}(0)=\delta_{n,0}$ (Fig.2) and $c_{n}(0)=\delta_{n,10}$ (Fig.3),  for both a uniform and semi-infinite lattice  $\kappa_n= \rho$ [panels (a)] and for  a Glauber-Fock lattice $\kappa_n=\rho \sqrt{n}$ [panels (b)]. The driving force is taken to be sinusoidal $F(t)=F_0 \cos (\omega t)$; parameter values are $\omega/ \rho=0.5$ and $F_0/ \omega =2.405$, corresponding to the first root of $J_0$ Bessel function. The figures clearly show that self-imaging after each oscillations cycle is achieved solely in the Glauber-Fock lattice case. In particular, in the uniform lattice self-imaging fails owing to lattice truncation.\\  
An experimentally-accessible model system to demonstrate DL and self-imaging in ac-driven Glauber-Fock lattice could be provided by light transport in an array of evanescently-coupled optical waveguides with a sinusoidally-curved axis (see, for instance \cite{Garanovich2,GaranovichReview}). Discretized light in waveguide optical lattices has provided in the past recent years a test bed to simulate in optical settings coherent quantum transport phenomena in the matter  \cite{GaranovichReview,LonghiReview}. To realize Glauber-Fock lattices, a chain of optical waveguides can be arranged in the geometrical setting of Fig.1(b). Adjacent waveguides in the chain are spaced by the same distance $d$ along the $x$ axis, whereas a decreasing distance along the $y$ axis is applied in such a way that the coupling constant between adjacent waveguides increases with site index $n$ according to Eq.(3) (see, e.g. \cite{LonghiPRBR}). The optical axis of the waveguides is then sinusoidally-bent in the $(x,z)$ plane along the spatial propagation distance $z$ with a spatial frequency $ \omega= 2 \pi /T$ to mimic the effect of the external sinusoidal force \cite{GaranovichReview,LonghiReview,SzamJPB}. Coupled-mode equations describing light transport in this waveguide chain reproduce the inhomogeneous lattice model defined by Eq.(3). Design parameters for the waveguide lattice are similar to those suggested, for instance, in Ref.\cite{LonghiPRBR} to study Bloch oscillations and metal-insulator transitions in inhomogeneous lattices, and should be accessible with the current femtosecond laser writing technology \cite{SzamJPB}. We note that in an experimental realization the ideal Glauber-Fock lattice model is realized only for a  limited number of sites, since the array is truncated at same site $n=N$.  The system dynamics is thus correctly reproduced in the photonic model system provided that the occupation amplitudes of sites $n>N$ is negligible. For example, in the simulations shown in Figs.2(a) and (b), the minimum number $N$ of the waveguides is $N=30$, which is feasible with current experimental set ups \cite{SzamOL}.  

\section{Conclusions}

In conclusion,  we have theoretically shown that dynamic localization and quantum-self-imaging can be realized in a class of inhomogeneous tight-binding lattices, the so-called Glauber-Fock lattices. Such a class of lattices have been recently introduced to simulate classical analogues of displaced Fock states, and is encountered in the Fock space representation of certain solid-state models or simple quantum field theoretical models. The self-imaging property  
of Glauber-Fock lattices induced by an ac force is a rather singular result, because inhomogeneous lattices are unlikely to show  exact quasi energy band collapse \cite{Longhi08}.  An experimentally accessible system to visualize dynamic localization and the self-imaging property of Glauber-Fock lattices could be provided by a photonic model system based on light transport in a chain of sinusoidally-curved optical waveguides with engineered coupling constants.  \par

This work was partially supported by the Italian MIUR (Grant No. PRIN-20082YCAAK). S.L. acknowledges hospitality at the IFISC (CSIC-UIB), Palma de Mallorca.


\begin{thebibliography}{99}
\bibitem{note0}
The case of a time-varying hopping rate $\rho=\rho(t) \geq 0$ can be always reduced to the case of a constant hopping rate $\rho=1$ and a time-varying (re-normalized) force. In fact, let us assume the general case $\rho=\rho (t) \geq 0$ in Eq.(2). Then, after introduction of the new time variable $\tau= \int_0^t dt' \rho(t')$, it readily follows that Eqs.(1) take the form
$ i (dc_n / d \tau)=-\sqrt{n} c_{n-1}-\sqrt{n+1} c_{n+1}+n f (\tau) c_n$, where the re-normalized force $f(\tau)$ is given by $f(\tau)=F(t(\tau)) / \rho(t(\tau))$ and $t(\tau)$ is obtained as the inverse function of $\tau(t)=\int_0^t dt' \rho(t')$. Therefore, we can limit our analysis by assuming a constant value of $\rho$ in Eq.(2). Note that our analysis includes, as a special case,  the constant force $F(t)=F_0$ oscillator case with a time-modulated hopping rate previously investigated in Ref. \cite{SzamGFb}.
\bibitem{note}
In the high-frequency limit $\rho / \omega \rightarrow 0$, DL is approximately realized for any inhomogeneity $\kappa_n$ of hopping rates and basically corresponds to coherent destruction of tunneling between adjacent lattice sites. In fact, in this case application of the rotating-wave approximation to Eqs.(3) yields an effective renormalization of the hopping rates $\kappa_n(t) \rightarrow \kappa_n (1/T) \int_0^T dt \exp[-i\Phi(t)]$, which vanish when the DL condition (16) is satisfied.  However, this result is only an approximate one and holds up to the time scale $ t \sim T( \omega / \kappa_0)$: tunneling is generally not suppressed at longer time scales (see \cite{Longhi08}).
\end{thebibliography}
\end{document}